\documentclass[prl,aps,twocolumn,superscriptaddress]{revtex4-1}
\usepackage{graphicx,ifpdf}
\usepackage[usenames]{color}
\usepackage{amsmath,amssymb}
\usepackage{gensymb}
\usepackage{natbib}
\usepackage{xcolor}
\def\strutdepth{\dp\strutbox}
\def\nw#1{\strut\vadjust{\kern-\strutdepth\vtop to0pt{\vss\hbox to\hsize
{\hskip\hsize\hskip5pt$\leftarrow$\hss\strut}}}{\em #1}}

\ifpdf 
\usepackage[pdftitle={Moving contact line of a volatile fluid},
colorlinks,linktocpage,breaklinks,pdftex,hyperfigures]{hyperref} \fi

\begin{document}

\title{Moving contact line of a volatile fluid}
\author{V. Jane\v{c}ek}
\affiliation{Physique et M\'ecanique des Milieux H\'et\'erog\`enes, UMR 7636 ESPCI -- CNRS -- Univ.~Paris-Diderot -- Univ.~P.M.~Curie, 10 rue Vauquelin, 75005 Paris, France}
\affiliation{ESEME, Service des Basses Temp\'eratures, UMR-E CEA / UJF-Grenoble 1, INAC, Grenoble, France}
\author{B. Andreotti}
\affiliation{Physique et M\'ecanique des Milieux H\'et\'erog\`enes, UMR 7636 ESPCI -- CNRS -- Univ.~Paris-Diderot -- Univ.~P.M.~Curie, 10 rue Vauquelin, 75005 Paris, France}
\author{D. Pra\v{z}\'{a}k}
\author{T. B\'{a}rta}
\affiliation{Department of Mathematical Analysis, Charles University, Sokolovska 83, 186 75, Prague, Czech Republic}
\author{V. S. Nikolayev}

\affiliation{Physique et M\'ecanique des Milieux H\'et\'erog\`enes, UMR 7636 ESPCI -- CNRS -- Univ.~Paris-Diderot -- Univ.~P.M.~Curie, 10 rue Vauquelin, 75005 Paris, France}
\affiliation{ESEME, Service des Basses Temp\'eratures, UMR-E CEA / UJF-Grenoble 1, INAC, Grenoble, France}

\begin{abstract}
Interfacial flows close to a moving contact line are inherently multi-scale. The shape of the interface and the flow at meso- and macroscopic scales inherit an apparent interface slope and a regularization length, both called after Voinov, from the dynamical processes at work at the microscopic level. Here, we solve this inner problem in the case of a volatile fluid at equilibrium with its vapor. The evaporative/condensation flux is then controlled by the dependence of the saturation temperature on interface curvature --~the so-called Kelvin effect. We derive the dependencies of the Voinov angle and of the Voinov length as functions of the substrate temperature. The relevance of the predictions for experimental problems is finally discussed.
\end{abstract}

 \date{\today}

\maketitle

\begin{figure}[t]
\includegraphics{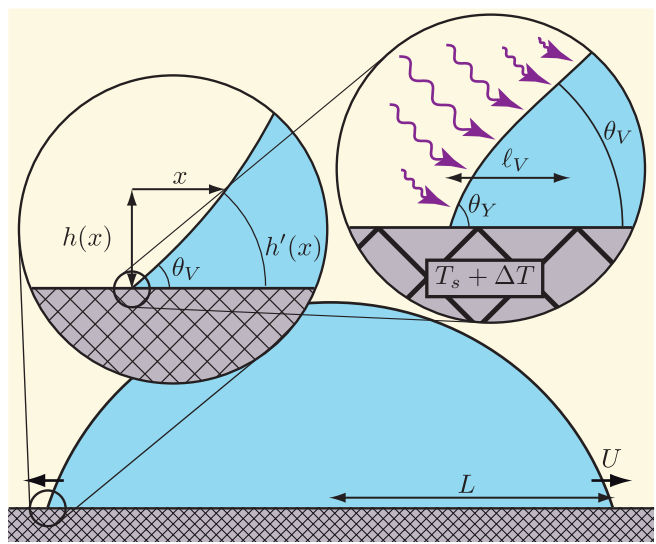}
\vspace{-2 mm}
\caption{(Color online) Schematic showing a liquid vapor interface $h(x)$ --~here, a spreading drop ($U>0$) on a cooled plate ($\Delta T<0$)~-- at different scales. The inner region, close to the moving contact line, is controlled by evaporation/condensation (top right). The slope changes from the Young angle $\theta_Y$ to the Voinov angle $\theta_V$ across a scale $\ell_V$ given by the Kelvin length $\ell_K$. In a mesoscopic range of scales, the shape of the interface results from the balance between viscous friction and Laplace pressure gradient, resulting in a slope $h'(x)$ which varies logarithmically in scale (Eq.~\ref{VoinovLaw}).}
\vspace{-2 mm}
\label{Schematic}
\end{figure}

The dynamics of a macroscopic solid plunging in a liquid bath \cite{DYCB07,MCSA12} or withdrawn from it \cite{E04b,E05,SDAF06} depends sensitively on its wetting properties i.e. on the intermolecular interactions at the nanoscopic scale. The motion of the contact line separating wet from dry regions is therefore an inherently multi-scale problem. Amongst the important consequences of the coupling between inner and outer scales (Fig.~\ref{Schematic}),  the speed at which a contact line can recede over a flat solid surface cannot exceed a critical value, associated to a dynamical wetting transition which leads to the formation of a dewetting ridge \cite{RBR91,SnE10}, of a V-shaped dewetting corner \cite{BR79,PSDL09,DYCB07,DFSA07} or to the entrainment of films \cite{DFSA07,MCSA12,SP91} (see \cite{BEIMR09,SA13} for detailed reviews). In many applications, such as coating, imbibition of powders, immersion lithography or boiling-free heating, these entrainment phenomena are crucial limiting factors for industrial processes.

Fig.~\ref{Schematic} shows schematically the structure of the flow close to a moving contact line. Even for an infinitesimal velocity $U$, there exists a range of mesoscopic scales --~roughly six decades~-- separating the microscopic scale from the macroscopic length $L$, in which the diverging viscous stress is balanced by a gradient of capillary pressure.  This balance can be made quantitative in the lubrication approximation, for which the angles are assumed small, and which gives a third order differential equation for the interface profile $h(x)$:
\begin{equation}\label{eq:lubrication}
\gamma \frac{d^3h}{dx^3} = - \frac{ 3\eta U}{h^2}.
\end{equation}
where $\eta$ is the liquid dynamic viscosity and $\gamma$ the surface tension; $U$ is positive for an advancing contact line. This equation has an exact solution \cite{DW97} which reduces to the asymptotic form proposed by Voinov~\cite{V76} far from the contact line, but for $x\ll L$:
\begin{equation}\label{VoinovLaw}
h'(x)^3=\theta_V^3+\frac{9\eta U}{\gamma}\, \ln\left(\frac{x}{\ell_V}\right).
\end{equation}
$\theta_V$ is by definition the apparent contact angle in the static case ($U=0$), which can be different from the Young angle $\theta_Y$ due to out of equilibrium processes taking place at a microscopic scale. The Voinov length $\ell_V$ is also a quantity defined in the mesoscopic range of scales but inherited from the inner region, where the problem is regularized. The mesoscopic solution (\ref{VoinovLaw}) must also be matched at the macroscopic scale $L$ to an outer solution where viscosity can usually be neglected. Fig.~\ref{Schematic} features the case of a spreading drop or of a growing bubble but the outer matching problem has been solved for many other geometries including a gravity controlled bath \cite{CSE12,E04b,E05} or a capillary ridge \cite{SnE10}.

Different models have been proposed to solve the moving contact line paradox i.e. the singularity of Eq.~(\ref{eq:lubrication})  as $x\to 0$. The simplest regularization is obtained by imposing the Navier slip boundary condition based on a slip length $\ell_s$ that can be expressed using a statistical physics description of liquids \cite{HSHNB08} and gas \cite{Bo93}. The Voinov angle is then Young's angle $\theta_Y$ and the Voinov length reads $\ell_V=3\ell_s/(e\;\theta_V)$, where $e$ is Euler's number \cite{E05}. Alternative descriptions have been proposed, based on disjoining pressure ($\ell_V$ then scales on the Israelachvili's length $(\mathcal{A}/6\pi\gamma)^{1/2}$, where $\mathcal{A}$ is the Hamaker constant)  and diffuse interface models ($\ell_V$ is then set by a diffusion length). Finally, when the substrate present heterogeneities, the contact line dynamics becomes in the inner layer a thermally activated process~\cite{Rolley,SA13}.

Here we consider the contact line motion of a volatile liquid in contact with an atmosphere of its pure vapor, see \cite{Ajaev10,PF10,Rednikov11,EuLet12,Kunkelmann12} and references therein. We show that the lubrication equation is perfectly regular when evaporation/condensation processes are taken into account. The Voinov length $\ell_V$ and the Voinov angle $\theta_V$ as functions of the substrate temperature constitute the central results of this Letter. The relevance of the theory for the case of a drop evaporating in air, usually assumed to be controlled by vapor diffusion \cite{deegan,eggers10,GBS12,PBLL10}, will be discussed in the conclusion.

\textit{Lubrication equations including Kelvin effect~--~}Evaporation/condensation process has first been proposed as a possible mechanism controlling the contact line motion at the molecular scale by Wayner \cite{Wayner93} and Pomeau \cite{P00}. In this Letter, we formalize this idea in a hydrodynamical framework, starting from the equation governing the evolution of the interface position $h$:
\begin{equation}\label{massconservation}
\partial_t h+\partial_x q=-j
\end{equation}
where $q$ is the hydrodynamic flow rate. The rate $j$ at which a liquid evaporates is governed by the energy balance at the liquid-gas interface. Assuming that the vapor pressure is fixed, the interfacial temperature $T^i$ depends on the interface curvature $\kappa\simeq h''(x)$ according to Kelvin's law
\begin{equation}\label{Ti}
T^i=T_s\left(1+\frac{\gamma \kappa}{\rho \mathcal{L}}\right),
\end{equation}
where $T_s$ is the saturation temperature, $\mathcal{L}$ the latent heat and $\rho$ the liquid density. In the lubrication approximation, the temperature varies linearly across the liquid layer from the substrate temperature $T_s+\Delta T$, assumed to be imposed (Fig.~\ref{Schematic}), to the interfacial temperature $T_i$. Neglecting the energy flux in the vapor phase, the evaporation rate is controlled by the conductive energy flux across the liquid,
\begin{equation}\label{j}
j=\frac{k}{\rho \mathcal{L}h}\left(\Delta T-\frac{T_s\;\gamma \kappa}{\rho \mathcal{L}}\right).
\end{equation}
where $k$ is the liquid heat conductivity.
\begin{figure}[t!]
\includegraphics{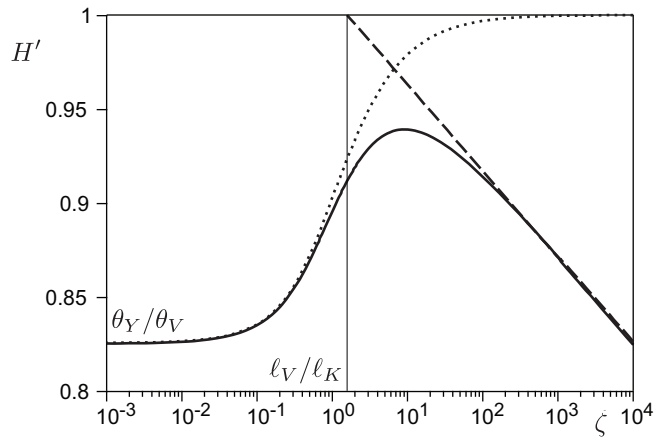}
\vspace{-2 mm}
\caption{Solution of the equations for a receding contact line ($\delta=-0.02$), with an overheating $\epsilon=0.1$. The dotted line corresponds to the static case ($\delta=0$). The dashed line corresponds to the Voinov outer asymptotics (\ref{VoinovLaw}). These solutions allow one to obtain the Voinov length $\ell_V$ and angle $\theta_V$.}
\label{Solution}
\vspace{-2 mm}
\end{figure}

Starting from the Voinov law (\ref{VoinovLaw}),  we define the reduced capillary number \cite{Eggers05}, using $\theta_V$ as a characteristic slope:
\begin{equation}
\delta \equiv \frac{3\eta U}{\gamma \theta_V^3}.
\end{equation}
The characteristic length $\ell_K$ is obtained dimensionally by balancing the two fluxes driven by the interface curvature, namely the evaporation rate $j$ and the divergence of the hydrodynamical flow rate $q=\gamma h^3 \kappa'/(3\eta)$:
\begin{equation}
\ell_K \equiv \frac{\sqrt{3 \eta k T_s}}{\theta_V^2 \rho \mathcal{L}}.
\end{equation}
We therefore make the solution dimensionless using
\begin{equation*}
h(x)=\theta_V\ell_K H(\zeta)\;, \quad \zeta=x /\ell_K,
\end{equation*}
Under the lubrication approximation, the governing equations in the scaled variables then read
\begin{equation}\label{leqf}
H''={\mathcal K},\quad  {\mathcal K}'=QH^{-3 }-\delta H^{-2},\quad Q'=({\mathcal K}-\epsilon)/H,
\end{equation}
where $\epsilon$ is the superheating parameter, defined by
\begin{equation}
\epsilon\equiv\frac{\sqrt{3 \eta k T_s}}{\gamma \theta_V^3}\;\frac{\Delta T}{T_s}
\end{equation}
and $Q$ is the dimensionless counterpart of $(q+Uh)$. The fourth order differential equation \eqref{leqf} must be complemented by the appropriate boundary conditions. We choose $x=0$ for the contact line position (so $H(0)=0$) and impose the slope $H'(0)=\theta_Y/\theta_V$ according to Young's law. As we look for regular solutions of the problem, the continuity of the temperature at the contact line requires ${\mathcal K}(0)=\epsilon$. Finally, to make the problem compatible with the asymptotic expansion \eqref{VoinovLaw}, one assumes a vanishing curvature far from the contact line: ${\mathcal K}(\infty)\to0$.

\begin{figure}[t]
\includegraphics{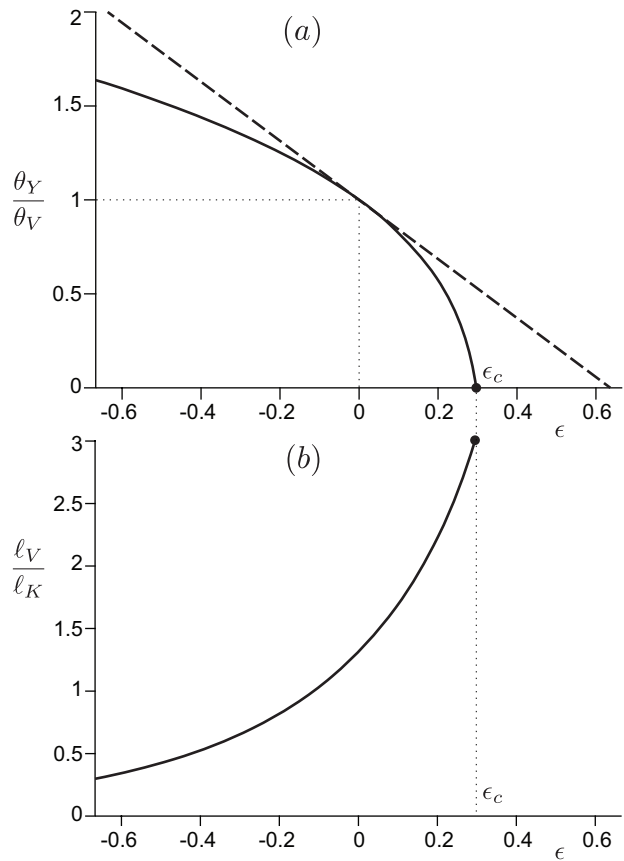}
\vspace{-2 mm}
\caption{(a) Ratio of Young angle to Voinov angle as a function of overheating parameter $\epsilon$, determined numerically (solid line). The dashed line is the analytical expansion (\ref{Appr1}).}
\vspace{-2 mm}
\label{LVoinov}
\end{figure}

\textit{Voinov angle~--~}By definition, $\theta_V$ is the interface slope for $x\gg \ell_K$ at vanishing capillary number $\delta$. In this limit, the outer boundary condition ${\mathcal K}(\infty)\to0$ is equivalent to the constant slope condition $H'(\infty)=1$. The dotted line in Fig.~\ref{Solution} corresponds to a typical solution obtained numerically for $\delta=0$  \cite{Suppl}. Overheating ($\epsilon>0$) induces an evaporation flux that would diverge as $h^{-1}$ at the contact line (cf. Eq. \ref{j}), if not balanced by Kelvin's effect. The induced liquid flow towards the contact line is accompanied by a capillary pressure gradient: the resulting interface curvature leads to an apparent angle $\theta_V$ larger than $\theta_Y$. The cross-over from $\theta_Y$ to $\theta_V$ takes place at the microscopic scale, for $\zeta$ of the order unity. The ratio $\theta_Y/\theta_V$ is reported in Fig.~\ref{LVoinov}a as a function of $\epsilon$. Following \cite{EuLet12}, we perform a linear expansion of the solution in $\epsilon$ for $\delta=0$, writing $H\equiv H_0=\zeta+\epsilon H_\epsilon+\mathcal{O}(\epsilon^2)$. Linearizing Eq.~\ref{leqf}, one obtains a differential equation on ${\mathcal K}_\epsilon=H_\epsilon''$, which reads $(\zeta^3 {\mathcal K}_\epsilon')'-{\mathcal K}_\epsilon/\zeta=-1/\zeta$. The solution verifying the boundary conditions involves the modified Bessel function of the first order $K_1$: ${\mathcal K}_\epsilon=1-\zeta^{-1} K_1\left(\zeta^{-1}\right)$. Integrating ${\mathcal K}_\epsilon$ from $\infty$ to 0, one obtains $H_\epsilon'(0)=-\pi/2$, which gives the expansion for the Voinov angle
\begin{equation}\label{Appr1}
\theta_Y/\theta_V= 1-(\pi/2)\;\epsilon+\mathcal{O}(\epsilon^2),
\end{equation}
shown in dashed line in Fig.~\ref{LVoinov}a.

The most important feature of the curve $\theta_Y/\theta_V(\epsilon)$ is the existence of a critical value $\epsilon_c\simeq 0.297$ of the overheating parameter --~note that the linear approximation (\ref{Appr1}) overestimates $\epsilon_c$ by a factor $\simeq 2$. In the limit $\epsilon\to\epsilon_c$, $\theta_V$ becomes much larger than $\theta_Y$. Then, Kelvin effect is just balanced by the maximal available capillary force, so that $H'$ goes from $0$ to $1$ \cite{Suppl}. The equation $\epsilon=\epsilon_c$ gives the large $\Delta T$ asymptotic expression of the Voinov angle $\theta_V$, which does not depend any longer on $\theta_Y$:
\begin{equation}
\theta_V \approx \left(\frac{\sqrt{3 \eta k T_s}}{ \epsilon_c \gamma}\;\frac{\Delta T}{T_s}\right)^{1/3}\;.
\end{equation}
One may expect this asymptotic regime to be relevant close to the gas-liquid critical point, in particular to the description of boiling \cite{PRE01,PRL06}.

\textit{Voinov length~--~}We now consider a contact line moving at a velocity $U$. We linearize the governing equations with respect to $\delta$, around the solution $H_0$ obtained for $\delta=0$ \cite{Suppl}. A typical solution is shown in Fig.~\ref{Solution} (solid line) together with the asymptotic expansions around $\zeta\to0$ (static solution obtained for $\delta=0$, dotted line) and $\zeta\to\infty$ (Voinov expansion, dashed line). It shows that a perfectly regular solution is obtained, in spite of the no-slip boundary condition imposed at the solid-liquid interface: at a scale smaller than $\ell_K$, the interface advances by the curvature driven condensation (or recedes by evaporation). How can a contact line advance (even at $\epsilon=0$) in the absence of any regularizing mechanism leading to a slip of the fluid at the boundary? Consider a perfect wedge initially at rest at the Young angle. Imposing an hydrodynamics flux towards the contact line leads to an increase of the apparent contact angle. However, $\theta_Y$ remains to be the true contact angle at the molecular scale so that a positive curvature $\kappa$ appears at a small scale, which induces a condensation by Kelvin effect. As a consequence, the contact line advances although the liquid velocity vanishes at the contact line: the phase transition flux $j(0)=-U\theta_Y$ balances exactly that induced by the contact line motion.

At distances much larger than the scale $\ell_V$, one recovers as expected the Voinov solution $H'(\zeta)\sim 1+\delta \log(\zeta\ell_K/\ell_V)+{\mathcal O}(\delta^2)$. The Voinov length $\ell_V$ is obtained from the matching to this outer expansion, as shown geometrically in Fig.~\ref{Solution} (intersection between the dashed line and the horizontal line $H'=1$). Fig.~\ref{LVoinov} shows the dependence of $\ell_V$ on the overheating parameter $\epsilon$. As expected from dimensional analysis, $\ell_V$ is on the order of the Kelvin length $\ell_K$. The ratio $\ell_V/\ell_K$ turns out to increase with the overheating parameter, from $\simeq 1.32$ at $\epsilon=0$ to $\simeq 3.00$ at $\epsilon=\epsilon_c$.

\textit{Concluding remarks~--~}As long as one aims to address a macroscopic problem involving a moving contact line, the only quantities inherited from the inner molecular-scale region are the Voinov length and the Voinov angle. For a volatile fluid with a vanishing slip length, we have shown here that the Voinov length is set by the Kelvin length. However, a true fluid presents both slip at the solid/liquid interface and evaporation/condensation at the solid/vapor interface. The theory developed in this Letter is applicable if the Voinov length produced by Kelvin effect is larger than that induced by the slip length i.e. if the condition $\theta_V\;\ell_K/\ell_s>1$ is fulfilled. The product $\theta_V^2\;\ell_K$ depends only on the liquid properties and ranges from $0.3$~nm for water and methanol to $1$~nm for alkanes and refrigerants such as ammonia or fluorocarbon. It can be even larger, for fluids like glycerol or silicon oils whose large viscosities are due to glassy effects. As the slip length $\ell_s$ is around two molecular sizes when $\theta_Y<\pi/2$, a good proxy of $\theta_V\;\ell_K/\ell_s$ is $\theta_V^{-1}$: evaporation is indeed the regularizing mechanism for any contact line problem, in the limit of low contact angles.

The theory developed here is directly applicable to the microscopic description of boiling, as nucleating bubbles are constituted of pure vapor. It may help resolving the demanding problem of the boiling crisis \cite{PRL06}. However, a second evaporative effect has been neglected here: when a fluid molecule joins the vapor phase, its momentum is lost by the liquid. A normal stress $\rho^2 j^2/\rho_v$ is therefore exerted on the interface ($\rho_v\ll\rho$ is the vapor density). Balancing this effect with capillary pressure $\gamma/l_K$, one deduces that the vapor recoil effect is dominant when the contact line velocity $U$ is much larger than the recoil velocity
$U_r= ( \gamma \mathcal{L})^{1/2}(\rho_v/\rho)^{1/2}\;(3 \eta k T_s)^{-1/4}$. $U_r$ is very large ($\sim 10$ m/s for a fluorocarbon) so that vapor recoil pressure can usually be neglected.

Is Kelvin's effect relevant to the coffee stain problem \cite{deegan} where a drop evaporates in air? The rate of evaporation $j$ in such a case must satisfy simultaneously three conditions: (i) the kinetic equation based on Hertz-Knudsen law, (ii) the conservation of mass, controlled by the diffusion of vapor and (iii) the energy conservation at the interface. The problem reduces to that considered here if the first two processes are fast enough, i.e. if the evaporation rate predicted by Eq.~\eqref{j} is the limiting factor. (i) The kinetics of evaporation involves thermal velocities of a fraction of the speed of sound in air, much larger than observed evaporation rates.  (ii) The evaporation rate $j$ predicted by the diffusion alone \cite{deegan,eggers10,GBS12,PBLL10} diverges at the contact line as $x^{-1/2}$. (iii) In conclusion, the evaporation rate at the microscopic scale must be generically controlled by the energy brought by thermal conduction to the liquid/air interface. Kelvin's effect shall therefore regularize both the stress and the evaporative flux singularities induced by the wedge geometry of a contact line \cite{ColinetArxiv12}.

\end{document}